\documentclass[12pt]{article}
\usepackage{verbatim,graphicx,amssymb,cite}
\newcommand{\be}{\begin{equation}}
\newcommand{\ee}{\end{equation}}

\begin{document}
\setcounter{page}{0}
\begin{titlepage}
\vspace*{0.1cm}
\begin{center}
{\Large \bf Long baseline neutrino experiments, mass hierarchy and $\delta_{CP}$ 
}\\
\vspace{1.0cm}

{\large
C. R. Das\footnote{E-mail: crdas@cftp.ist.utl.pt},
Jo\~{a}o Pulido\footnote{E-mail: pulido@cftp.ist.utl.pt}\\
\vspace{0.15cm}
{{\small \sl CENTRO DE F\'{I}SICA TE\'{O}RICA DE PART\'{I}CULAS (CFTP)\\
 Departamento de F\'\i sica, Instituto Superior T\'ecnico\\
Av. Rovisco Pais, P-1049-001 Lisboa, Portugal}\\
}}
\vspace{0.25cm}
\end{center}
\vglue 0.6truecm

\begin{abstract}
We investigate the possibilities offered by the long baseline experiments T2K, No$\nu$a, 
LBNE and LAGUNA for the evaluation of the neutrino mass hierarchy and CP violating phase 
$\delta_{CP}$. We consider a neutrino and antineutrino energy in the interval [0.5,12] GeV.
It is found that the clearest possible distinction between the two hierarchy signatures is
provided by LAGUNA for an (anti)neutrino energy $E_{(\bar\nu)\nu}\simeq 0.5$ GeV in the $(\bar\nu_{\mu}
\rightarrow \bar\nu_{\mu})~\nu_{\mu}\rightarrow \nu_{\mu}$ disappearance channel. For LBNE at 
$E_{\bar\nu}\simeq 1$ GeV the $\bar\nu_{\mu}\rightarrow \bar\nu_{\mu}$ channel may also provide 
a distinction, although not so clear, and for No$\nu$a this may be even less clear. These results 
are essentially the same for the $\theta_{23}$ first and second octant solutions. As for 
$\delta_{CP}$ determination, LAGUNA also offers the best chances at $E_{\bar\nu}\simeq 0.5$ 
GeV in the $\bar\nu_{\mu}\rightarrow \bar\nu_{e}$ channel with $\theta_{23}$ 
in either octant. Regarding nonstandard interactions, the best possibility for their 
investigation resides in No$\nu$a whose source-far detector distance can become a magic 
baseline if the hierarchy is normal for channels $\nu_{\mu}\rightarrow \nu_{\mu}$,
$\bar\nu_{\mu}\rightarrow \bar\nu_{e}$, $\bar\nu_{\mu}\rightarrow \bar\nu_{\mu}$ with
energy (8.9-9.1) GeV, (11.6-12) GeV, (8.8-12) GeV. If the hierarchy is inverse, magic
baselines for No$\nu$a occur for $\nu_{\mu}\rightarrow \nu_{e}$, $\nu_{\mu}\rightarrow \nu_{\mu}$,
and $\bar\nu_{\mu}\rightarrow \bar\nu_{\mu}$ with energy (11.8-12) GeV, (8.9-12.3) GeV,  
(8.8-8.9) GeV. For LAGUNA a magic baseline appears for the $\bar\nu_{\mu}\rightarrow \bar\nu_{\mu}$
channel only at (4.7-5.0) GeV in inverse hierarchy. We have also investigated a possible
complementarity between T2K and No$\nu$a.
\end{abstract}

\end{titlepage}

\section{Introduction}

The solar and atmospheric neutrino anomalies \cite{Cleveland:1998nv} 
-\cite{Wendell:2010md} have since long ago motivated interest \cite{Parke:1993dp} in 
long baseline (LBL) neutrino experiments (\cite{Barish:1999zz} - \cite{Sirignano:2011zz}),
aiming at an accurate evaluation of some of the neutrino parameters \footnote {For a 
recent review on LBL neutrino experiments see\cite{Feldman:2012qt}.}. Among
these, the values of the mass squared differences and mixing angles are by now reasonably well 
established from the data of the solar \cite{Cleveland:1998nv} -\cite{Aharmim:2008kc}, atmospheric 
\cite{Wendell:2010md} and reactor experiments \cite{An:2012eh} -\cite{Abe:2011fz}. Still, 
several issues on neutrino parameters are still open, namely the absolute mass, the mass
hierarchy and the magnitude of $\delta_{CP}$, the $CP$ violating phase. LBL neutrino 
experiments, either from accelerators \cite{Adamson:2012rm} -\cite{Sirignano:2011zz},\cite{LBNE}, 
neutrino factories or beta beams \cite{Albright:2004iw},\cite{BetaBeams} will be essential in 
providing the answer to the two most important questions, namely, whether the hierarchy is normal 
(NH, $\Delta m^2_{31}>0$) or inverse (IH, $\Delta m^2_{31}<0$) and whether or not $CP$ 
violation occurs in the leptonic sector ($sin~\delta_{CP}\neq 0$). 

The recent measurement of the $\theta_{13}$ mixing angle \cite{An:2012eh},\cite{Abe:2011fz}
reduces the parameter degeneracy \cite{Barger:2001yr} -\cite{Ishitsuka:2005qi} inherent in the 
LBL three neutrino analysis, so that the above goals have become within reach of the forthcoming 
experiments. Besides the existing T2K \cite{Abe:2011ks} and No$\nu$a \cite{Nova}, the latter 
expected to start taking data soon, two other major projects are being considered: LBNE 
\cite{LBNE} and LAGUNA/LBNO \cite{Bertolucci:2012fb}. The first, whose proposal is under current 
evaluation, aims at sending a neutrino beam from Fermilab to Homestake ($\sim$ 1290 km) and the 
second favours a baseline from CERN to the Pyhas\"{a}lmi mine in Finland ($\sim$ 2290 km) with 
a proposal expected to be submitted by 2014 \cite{Bertolucci:2012fb}.


The present generation of LBL neutrino experiments \cite{Adamson:2012rm} -\cite{Sirignano:2011zz} 
relies on high energy neutrino beam facilities where charged pions and kaons are allowed to decay 
in flight. The neutrino (antineutrino) beam composition consists mostly 
of muon neutrinos (antineutrinos) with a small mixture of electron neutrinos (antineutrinos) 
and a small component of tau neutrinos if the beam energy is high enough. An alternative to 
obtain neutrino beams for LBL experiments has been proposed some time ago based on muon storage 
rings \cite{Geer:1997iz},\cite{Ayres:1999ug} whereby muons produced from meson decays are stored, 
accelerated and allowed to decay in the long straight sections of a storage ring \cite{Prior:2009zz}. 
These sections are aligned in the direction of the neutrino detectors which can be located several 
thousands of kilometers away. Since muons and antimuons decay to almost 100\% according to 
$$\mu^{-}\rightarrow e^{-}~\bar\nu_e~\nu_{\mu},~~~~\mu^{+}\rightarrow e^{+}~\nu_e~\bar\nu_{\mu},$$
the precise content of the neutrino and antineutrino beam can be accurately determined.

There have been many studies of $CP$ symmetries and mass hierarchy via neutrino oscillations for 
present and future facilities (see e.g.\cite{BurguetCastell:2002qx} -\cite{Agarwalla:2012bv})
and the prospects for their determination have considerably
improved after the measurement of $\theta_{13}$ \cite{An:2012eh},\cite{Abe:2011fz}. In
the present work we take advantage of this fact and explore the possibility for extracting 
$\delta_{CP}$ and the mass hierarchy from the T2K \cite{Abe:2011ks}, No$\nu$a 
\cite{Nova}, LBNE \cite{LBNE} and LAGUNA/LBNO \cite{Bertolucci:2012fb} LBL neutrino experiments. As 
shall be seen, by conveniently choosing a neutrino energy range and source-detector distance, the 
parameter degeneracy involving the mass hierarchy and $\delta_{CP}$ can be overcome with the data
from a single experiment. In fact to this end one detector located sufficiently far away 
from the source may be enough, as already noted in \cite{Winter:2008cn}.
Of course the possible redundancy provided by one or two extra detectors will always be welcome
for the verification and accuracy of the results
\footnote{The idea of two detectors placed at different baselines was first 
proposed in \cite{Minakata:1997td}. However, owing to the limited knowledge on the mass 
differences and mixing angles at the time, a solution to the degeneracies was far from reachable.}.

This paper is organized as follows: we start in section 2 with the derivation of the general 
expression for the oscillation probability in matter with constant density without resorting 
to any approximations. In section 3 we start by comparing the results of our numerical calculation
with the leading term approximations used in the literature \cite{Minakata:2001qm},
\cite{Cervera:2000kp},\cite{Akhmedov:2004ny}. From the oscillation probability expression 
we obtain the biprobability plots \cite{Minakata:2001qm} for a few relevant pairs of oscillation 
channels in which most of our analysis will be based. Thus in subsection 3.1 we get such plots 
for fixed distance, varying energy and several values of $\delta_{CP}$, in the cases of normal 
and inverse hierarchies. It will be seen that within preferred energy ranges the relative 
ordering of the iso-$\delta_{CP}$ curves remains invariant as the energy changes, while 
for other energy regions at the same distance they successively intersect each other.  
In these cases, for a small change in energy any two neighbouring curves may 
swop their relative position. This entails a less accurate, if at all possible, determination 
of $\delta_{CP}$ in the corresponding energy range at that distance. In fact the biprobability 
curves for constant energy as a function of $\delta_{CP}$ are closed contours which intersect the 
iso-$\delta_{CP}$ curves. Since the neutrino energy cannot be known with arbitrarily good accuracy,
the two parameters (beam energy and distance from source) must be conveniently chosen to ensure 
that the iso-$\delta_{CP}$ curves do not intersect each other within the energy uncertainty range. 
Moreover the regions of sudden variations of the $CP$ phase along the closed contours should be 
avoided so that $\delta_{CP}$ could be determined with a reasonable accuracy.
In subsection 3.2 we obtain the biprobabilities for fixed energy, varying distance and several values
of $\delta_{CP}$ also for both hierarchies. Closed contours are now obtained for constant
distances as a function of the $CP$ phase and the same rule as before applies for the requirement
of its accurate determination. Thus far our analysis will only use the first octant solution for
the $\theta_{23}$ mixing angle and is mainly focused on the No$\nu$a, LBNE and LAGUNA
experiments. It turns out that the latter experiment appears to be particularly promising as
regards its contribution to providing definite conclusions for the mass hierarchy and the 
$\delta_{CP}$ range. At the end of section 3.2 we provide a comparison between the $\theta_{23}$ 
first and second octant solutions in the three experiments. Section 4 is dedicated to a comparison 
of T2K and No$\nu$a emphasizing the prospective distinction in these experiments between the 
signatures of mass hierarchies and between the signatures of the first and second octant solutions 
for $\theta_{23}$. Finally in section 5 we summarize the present work and draw our conclusions.

\section{The oscillation probability}

The starting point for our neutrino oscillation analysis is the standard matter Hamiltonian which
in the flavour basis reads
\be
H^{(f)}=U
\left(\begin{array}{ccc}0 & 0 & 0\\ 0 & \frac{\Delta m^2_{21}}{2E} & 0 \\
0 & 0 & \frac{\Delta m^2_{31}}{2E}\end{array}\right)U^{\dagger}+\left(\begin{array}{ccc}V_k & 0 & 0 \\
0 & 0 & 0 \\ 0 & 0 & 0 \end{array}\right)
\label{H}
\ee
where $\Delta m^2_{ji}=m^2_{j}-m^2_{i}$, 
$V_k=\sqrt{2}G_F N_{e_k}$, $G_F$ is the Fermi constant and $N_{ek}$ is the electron density in the 
$k^{th}$ layer. The latter can be expressed in terms of the mass density $\rho_k$, the electron 
number density $Y_k$ and the nucleon mass $m_N$ as
\begin{equation}
N_{ek}=\frac{Y_k \rho_k}{m_N}~.
\end{equation}
For the cases we study we need only to consider propagation within the Earth's crust, for which 
$\rho_k=3~g~cm^{-3}$ and $Y_k=1/2$. In eq.{(\ref H)} $U$ is the usual leptonic mixing matrix $PMNS$ 
relating the neutrino mass basis to the flavour basis\footnote{We use the Particle Data Group
notation \cite{PDG} for the $U$ matrix.}
\be
|\nu_{\alpha}\!>=U|\nu_{i}\!>
\ee
and in a like manner for future purpose one can define a unitary transformation relating 
the mass matter eigenstates to the flavour ones
\be
|\nu_{\alpha}\!>=U^{'}|\nu^{'}_{i}\!>.
\ee
Given the definition of $U$ (eq.(3)), denoting by $H^{(m)}$ the mass basis Hamiltonian operating
between mass eigenstates $|\nu_i\!>$ and $<\nu_j|$ and $H^{(f)}$ the flavour basis one
operating between flavour eigenstates  $|\nu_{\alpha}\!>$ and $<\nu_{\beta}|$, we have
\be
<\nu_j|H^{(m)}|\nu_i>=<\nu_{\beta}|U H^{(m)}U^{\dagger}|\nu_{\alpha}>=
<\nu_{\beta}|H^{(f)}|\nu_{\alpha}>
\ee 
hence
\be
H^{(f)}=UH^{(m)}U^{\dagger}.
\ee
Since $H^{(f)}$ and $H^{(m)}$ are related by a unitary transformation their eigenvalues are 
the same, therefore denoting by $V_f$ and $V$ respectively the matrices that diagonalize 
$H^{(f)}$ and $H^{(m)}$
\be
V_{f}^{\dagger}H^{(f)}V_f=V^{\dagger}H^{(m)}V=H_D
\ee
where $H_D$ is the diagonalized Hamiltonian. Using (6) one obtains the relation between 
matrices $V_f$ and $V$,
\be
V_f=UV
\ee
so that $UV$ also diagonalizes the flavour basis Hamiltonian as can be easily checked from eqs.(1) 
and (6). Moreover applying the definition of $U^{'}$ (eq.(4))
\be
<\nu_{\beta}|H^{(f)}|\nu_{\alpha}>=<\nu^{'}_j|(U^{'})^{\dagger}H^{(f)}U^{'}|\nu^{'}_i>
\ee
which shows that $U^{'}$ also diagonalizes $H^{f}$, thus $U^{'}=V_f=UV$.

Since the neutrino Hamiltonian (1) satisfies a Schr\"{o}dinger like equation
\begin{equation}
i\frac{d}{dx}\nu_{\alpha}=H_{\alpha \beta}^{(f)}{\nu_{\beta}}~~~~~(\alpha,\beta=e,\mu,\tau)~,
\end{equation}
a state produced as flavour $\alpha$ at the origin becomes, after
traveling a distance $L$
\begin{equation}
\nu_{\beta}(L)=S_{\beta \alpha}(L,0) \nu_{\alpha}(0)
\end{equation}
where $S_{\beta \alpha}(L,0)$ satisfies the same Schr\"{o}dinger equation. So in index notation 
\begin{equation}
S_{\beta \alpha}(L,0)=(U^{'})_{\beta i}exp\left[-i\int_0^L (U^{'\ast})_{\gamma i}(H^{(f)})_{\gamma \delta}
(U^{'})_{\delta j}dx\right](U^{'\ast})_{\alpha i}=(U^{'})_{\beta i}e^{-i\lambda_i L}(U^{'\ast})_{\alpha i}
\end{equation}
where $\lambda_i$'s are the mass matter eigenvalues. The oscillation probability  
\begin{equation}
P(\nu_{\alpha} \rightarrow \nu_{\beta}, L)=|S_{\beta\alpha}|^2
\end{equation}
is then evaluated through 
\begin{equation}
P(\nu_{\alpha} \rightarrow \nu_{\beta}, L)=(U^{'\ast})_{\beta i}e^{i\lambda_i L}(U^{'})_{\alpha i}
(U^{'})_{\beta j}e^{-i\lambda_j L}(U^{'\ast})_{\alpha j}~~~({\rm no~sum~in}~\alpha,\beta)
\label{basis}
\end{equation}
with the following neutrino parameter values \cite{Fogli:2012ua}, 
$$sin^2\theta_{12}=0.31,~sin^2\theta_{13}=2.4\times 10^{-2},~sin^2\theta_{23}=0.39,$$
\be
\Delta m^2_{21}=7.6\times 10^{-5}eV^2,~~\Delta m^2_{31}=2.4\times 10^{-3}eV^2.
\ee
For the sake of comparison we will present one example with the second octant solution for $\theta_{23}$
\cite{Tortola:2012te}
\begin{equation}
sin^2\theta_{23}=0.61.
\end{equation}
We will use equation (\ref{basis}) throughout the paper as the basis for our calculations.  

\section{No$\nu$a, LBNE, LAGUNA}

The channels to analyse in this section are the ones for which the three experiments (No$\nu$a, LBNE
and LAGUNA) are dedicated, namely the muon neutrino disappearance channel ($\nu_{\mu} \rightarrow \nu_{\mu}$), 
the inverse golden channel or electron neutrino appearance ($\nu_{\mu} \rightarrow \nu_{e}$) and their 
antineutrino counterparts for normal and inverse hierarchy. We denote these channels as 
$\mu\mu $, $\mu e$, their counterparts as $\bar\mu \bar\mu $, $\bar\mu\bar e$ respectively
and accordingly their oscillation probabilities as $P_{\mu\mu},P_{\mu e},P_{\bar\mu\bar\mu},
P_{\bar\mu \bar e}$. As mentioned in the introduction our main results are presented in terms of 
biprobability graphs. 

In the current literature the oscillation probabilities for long baselines are usually calculated
by resorting to approximations up to first or second order in 'small' parameters such as 
\cite{Richter:2000pu},\cite{Minakata:2001qm},\cite{Cervera:2000kp},\cite{Akhmedov:2004ny},
\cite{Gago:2009ij}
$$\frac{2\sqrt{2}G_{F}N_e E}{\Delta m^2_{31}},~\frac{\Delta m^2_{21}}{2E}L,~\frac{\Delta m^2_{21}}
{\Delta m^2_{31}},~\theta_{13},~\frac{\Delta m^2_{21}}{\Delta m^2_{31}}=\epsilon\simeq\theta_{13}.$$
In particular, as it is known nowadays, the approximation $\epsilon\simeq \theta_{13}$ is rather
innacurate. Here we will use instead the exact numerical expressions from (14) and show in fig.1 
the comparison between our results and those from the approximations used in the literature. In 
the four top panels of fig.1 we plot the oscillation probability $P_{\mu e} $ 
as a function of distance for eight values of $\delta_{CP}$ equally 
spaced from 0 to 360 degrees and in the two bottom panels the biprobability plots for the 
channel pair $\mu e,\bar\mu \bar e$. 
Here we consider normal hierarchy only and a neutrino energy E=2.3 GeV. Notice the large 
discrepancies in the probabilities and the locations of the intersections from panel to panel. 
These intersections correspond to the magic baselines to which we will return in section 3.2.

\subsection{Fixed distance biprobability plots}

In this subsection we just consider the No$\nu$a \cite{Nova} and LBNE \cite{LBNE} experiments 
and study the constant $\delta_{CP}$ curves as a function of energy for fixed distance.

No$\nu$a is planned to start taking data in 2013 and consists of a 330 ton near detector at the
Fermilab site and a 14 kiloton far detector. The latter is a liquid scintillator  
situated 12 km off-axis at 810 km in the Neutrino Main Injector (NuMI) beam produced at Fermilab. It 
is dedicated mainly to observe $\nu_{\mu} \rightarrow \nu_{e}$ oscillation along
with its antiparticle counterpart. The advantage of an off-axis location is that the neutrino energy 
is nearly independent from the parent meson energy, therefore the beam energy spread is much smaller 
than for an on-axis one \cite{Litchfield:2005zd}. Also using the NuMI beam, 
but not yet finally approved, is LBNE with a 34 kton liquid Argon far detector at 1290 km distance
and a neutrino energy in the interval [0.5,12] GeV.

In each of the four panels of fig.2 we show the biprobability curves for the pair of channels $\mu e$, 
$\mu\mu$ for eight values of $\delta_{CP}$ equally spaced from 0 to 360 degrees with energy running from 
0.5 GeV to 100 GeV. The top two panels refer to a distance of 810 km and the bottom two are for 1290 km 
which are the source-far detector distances for No$\nu$a and LBNE respectively. The left panels are for 
normal and the right ones are for inverse hierarchy. In each panel we plot the biprobability curves for 
three energies which are ellipses of large eccentricity: for 810 km we take
E=2.3 GeV which is the preferred No$\nu$a energy (middle ellipses in the top panels), E=1 GeV 
and 0.5 GeV (top and bottom ellipses respectively). In order to keep the figures as clear as 
possible, we choose not to continue the iso-$\delta_{CP}$ curves for E$<$1 GeV.

The most favourable energy range for the 
evaluation of $\delta_{CP}$ is the one where the iso-$\delta_{CP}$ curves lie the 
furthest apart since the phase variation is then the smoothest possible along each ellipse.
Consequently, as it is seen from all four panels of fig.2, a value near E=0.5 GeV is 
the best choice for $\delta_{CP}$ evaluation. For 1290 km 
(LBNE source-detector distance) we take E=7 GeV, 1 GeV and 0.5 GeV (top, middle and bottom
ellipses in the bottom panels of fig.2). Again we omit the iso-$\delta_{CP}$ curves
below 1 GeV and find that the best sensitivity to $\delta_{CP}$ is for 0.5 GeV, although a double
degeneracy remains in its determination due to the very large eccentricity of the ellipses. 
In the limit of increasing energies the constant $\delta_{CP}$ 
curves merge and eventually coincide at the point ($P_{\mu e}=0$, $P_{\mu \mu}=1$) whereas they diverge 
from each other for decreasing energy. This is consistent with the fact that the oscillation length 
increases for increasing energy, so for fixed distance the neutrinos oscillate less as their 
energy increases. The same as in fig.2 is done in figs.3 and 4 for the pairs of channels 
$\bar\mu\bar e$, $\bar\mu\bar\mu$ and $\mu e$, $\bar\mu \bar e$ respectively. In fig.3 
(for $\bar\mu\bar e$, $\bar\mu\bar\mu$) and in the top panels, the middle ellipses are for
E=2.3 GeV, the top and bottom ones for E=1 GeV and 0.5 GeV respectively, all for 810 km 
distance as in fig.2. In the bottom panels (1290 km) the top, middle and bottom ellipses 
are for E=7 GeV, 1 GeV and 0.5 GeV. Once again the iso-$\delta_{CP}$ curves are ommited 
below 1 GeV while 0.5 GeV is the most convenient choice as regards $\delta_{CP}$ evaluation,
if not for a double degeneracy. As for the channel pair $\mu e$, $\bar\mu \bar e$ (fig.4) 
the 0.5 GeV contours are again the ellipses with the longest major axes. So this energy is
also the most convenient choice for $\delta_{CP}$ evaluation. It will be seen in the 
next section that an improved situation can be found with data from LAGUNA.

The possibility of distinction between hierarchies appears realistic so long as the neutrino 
energy can be appropriately tuned, as can be inferred from the comparison between the two bottom 
panels of fig.3 ($\bar\mu\bar e$, $\bar\mu\bar\mu$ channels at the LBNE source-detector distance). 
In fact for E=1 GeV, which corresponds to the middle ellipse, it is seen that $P_{\bar\mu 
\bar\mu} \simeq 0.48$ for inverse hierarchy whereas for normal hierarchy $0.62\lesssim 
P_{\bar\mu \bar\mu}\lesssim 0.66$. Not so clear a distinction can be provided at the No$\nu$a 
distance for the same energy, as one can conclude from the comparison between the two top 
panels in figs.2 and 3.

In a like manner as figs.2, 3, 4, fig.5 shows the biprobability curves for the $\mu\mu$, 
$\bar\mu\bar\mu$ channel pair in normal hierarchy for 0.5 GeV, 1 GeV and 2.3 GeV at 810 km. 
In this case the iso-$\delta_{CP}$ curves are nearly superimposed, so that an unrealistically 
large experimental accuracy would be needed in order to obtain information on $\delta_{CP}$. 
The same situation occurs for inverse hierarchy for the same energies and distance
and for normal and inverse hierarchies at 1290 km with 0.5 GeV, 1 GeV and 7 GeV. Hence the 
$\mu\mu$, $\bar\mu\bar\mu$ channel combination is of little use, if any.

\subsection{Fixed energy biprobability plots}

In this subsection we extend our analysis to include the LAGUNA experiment and
study the constant $\delta_{CP}$ curves as a function of distance for
fixed neutrino energy from the production point up to 2290 km. This is the CERN-Pyhas\"{a}lmi 
mine distance where the LAGUNA detector is planned to be installed observing neutrinos produced
from CERN. We consider two neutrino energies: E=2.3 GeV and 0.5 GeV. Our results are plotted in 
figs.6, 7 (for 2.3 GeV) and 8, 9 (for 0.5 GeV). The top panels 
in fig.6 contain the biprobability curves for the $\mu e$, $\mu \mu$ channel pair, the 
middle ones for $\bar\mu \bar e$, $\bar\mu \bar\mu$ and the bottom ones for $\mu e$, 
$\bar\mu \bar e$. Left hand panels are for normal and right hand ones for inverse hierarchy
respectively. The iso-$\delta_{CP}$ curves are shown for eight values of $\delta_{CP}$
equally spaced from 0$^o$ to $360^o$ as a function of distance, all diverging from a
common point at zero distance. The constant distance contours are for 810 km
(No$\nu$a, dotted lines), 1290 km (LBNE, dashed lines) and 2290 km (LAGUNA, full lines). 
\footnote{The preferred neutrino energy for LAGUNA has not been decided yet. Here we
consider the same energies for all three experiments so that the corresponding closed contours 
appear in the same panel.}. Note that the No$\nu$a and LBNE contours correspond to the ones 
already shown in fig.2 for $\mu e$, $\mu\mu$ now superimposed on the iso-$\delta_{CP}$ curves 
for varying energy instead of varying distance as before. In the top right and middle left
panels of fig.6 ($\mu e$, $\mu {\mu}$ channel pair, inverse hierarchy and $\bar\mu \bar e$, 
$\bar\mu \bar\mu$ channel pair, normal hierachy) all curves intersect at one 
point. This corresponds to one of the channel probabilities being independent 
of $\delta_{CP}$ at one particular distance, the so called magic baseline. Such distance
can be evaluated by expanding the appearance probability 
\begin{equation}
P_{\mu e}=|(U^{'})_{e i}e^{-i\lambda_i L}(U^{'*})_{\mu i}|^2
\end{equation}
and separating the $\delta_{CP}$ dependent and independent terms. We obtain
\be
P_{\mu e}=2s_{12}c_{12}s_{23}c_{23}s_{13}c_{13}^2(As_{\delta}+Bc_{\delta})+ \delta_{CP}~{\rm 
independent}~{\rm terms}
\ee
with
\begin{eqnarray}
A\!\!&\!=\!\!&\!sin(\lambda_{1}-\lambda_{2})L-sin(\lambda_{1}-\lambda_{3})L+sin(\lambda_{2}-\lambda_{3})L \\
\nonumber
B\!\!&\!=\!\!&\!c_{2\theta_{12}}[1-cos(\lambda_{1}-\lambda_{2})L]-cos(\lambda_{1}-\lambda_{3})L+
cos(\lambda_{2}-\lambda_{3})L
\end{eqnarray}
where $c_{12}=cos_{\theta_{12}},...$ and $L$ is the distance from the source to the detector.
In order to ensure the independence of $P_{\mu e}$ on $\delta_{CP}$ one must require the 
vanishing of the quantity in brackets in (18). To this end we note that the three arguments of 
the sines and cosines in the expressions for $A$ and $B$ cannot simultaneously vanish for any 
possible baseline distance and moreover such condition would only be a sufficient one for 
$P_{\mu e}$ to become independent of $\delta_{CP}$. So one must impose 
\be
tan~\delta_{CP}=-~\frac{B}{A}
\ee
for any $\delta_{CP}$, which requires $A=B=0$. In this way for E=2.3 GeV one gets for the first 
three magic baselines for the $\mu e$ channel in inverse hierarchy $L_{magic}\simeq 1980~km,
~\simeq 3960~km$ and $~\simeq 5940~km$ (see fig.7). Hence the common point of all curves in the 
top right panel of fig.6 corresponds to the shortest one at 1980 $km$. Analogously, for the  
$\bar\mu \bar e$, $\bar\mu \bar\mu$ channel pair in normal hierarchy, the magic baselines
occur for $\bar\mu \bar e$, the shortest one being located at $L_{magic}\simeq 2008~km$, and 
the following ones at approximately $4020~km $ and $6030~km$. They can be easily determined 
in the same way as in the previous case with the replacements
\begin{equation}
U^{'}~\rightarrow~U^{'*}~~~~~~V~\rightarrow~-V.
\end{equation}
Magic baselines are particularly useful for the investigation of nonstandard interactions
\cite{Gago:2009ij},\cite{Das:2010sd},\cite{Coelho:2012bp}.

Fig.8 shows the biprobability contours for a neutrino energy E=0.5 GeV and the same channels 
as fig.6 at the No$\nu$a (dotted), LBNE (dashed) and LAGUNA (full) distances. For simplicity 
the iso-$\delta_{CP}$ curves are ommited from fig.8.

From the inspection of figs.6 and 8 (neutrino energies 2.3 GeV and 0.5 GeV respectively) 
one can infer the prospects for distinguishing between normal and inverse hierarchies. To 
this end one must compare the left and right panels of these figures. Starting with fig.6 
(middle panels) it is seen that for LBNE alone (bottom ellipse appearing as a line segment) 
normal hierarchy gives $0.012\lesssim P_{\bar\mu \bar e} \lesssim 0.034$, 
$P_{\bar\mu \bar \mu} \sim 0.06$ while inverse hierarchy gives $0.042\lesssim 
P_{\bar\mu \bar e} \lesssim 0.078$, $P_{\bar\mu \bar \mu}\sim 0.09$. Also for LBNE and
appearance probabilities only, the comparison of the bottom panels if fig.6 shows for normal 
hierarchy $0.042\lesssim P_{\mu e}\lesssim 0.078$, $0.012\lesssim P_{\bar\mu \bar e}\lesssim 
0.034$ while for inverse hierarchy such probabilities are interchanged. Another typical case 
can be observed from the top panels of fig.6 with the uppermost ellipses which correspond to 
the LAGUNA experiment. Here it is seen that $0.012\lesssim P_{\mu e} \lesssim 0.060$ and 
$P_{\mu e}\lesssim 0.020$ for normal and inverse hierarchy respectively, hence some 
overlap exists in the two probabilities. As for the $\mu \mu$ channel one has
$0.91\lesssim P_{\mu \mu}\simeq 0.95$ and $0.97\lesssim P_{\mu \mu}\lesssim 1$ for normal 
and inverse hierarchies, a difference which will be difficult to detect experimentally. A 
similar conclusion may be drawn from the inspection of the remainder of fig.6. Therefore the 
distinction between the two hierarchies appears to be too difficult to trace with a neutrino
energy 2.3 GeV.

In contrast, for E=0.5 GeV the distinction between signatures from each hierarchy becomes much
more plausible especially with the LAGUNA experiment. In fact from fig.8 and 
observing the LAGUNA contours, the two top panels tell us that for normal hierarchy $0.05
\lesssim P_{\mu e}\lesssim 0.14$, $0.37\lesssim P_{\mu\mu}\lesssim 0.40$, while for inverse 
hierarchy $0.02\lesssim P_{\mu e}\lesssim 0.22$, $0.032\lesssim P_{\mu\mu}\lesssim 0.073$. 
Similarly from the middle panels in normal hierarchy $0.01\lesssim P_{\bar\mu \bar e}
\lesssim 0.21$, $0.36\lesssim P_{\bar\mu\bar\mu}\lesssim 0.50$ while in inverse hierarchy 
$0.02\lesssim P_{\bar\mu \bar e}\lesssim 0.16$, $0.03\lesssim P_{\bar\mu\bar\mu}\lesssim 0.06$.
Thus it follows that the best possibilities for hierarchy distinction lie on the investigation 
of channels $\mu\mu$ and $\bar\mu \bar\mu$ at an energy near 0.5 GeV in the LAGUNA experiment. 
In fact in all cases a dedicated observation of fig.8 shows that the difference between the 
probabilities for normal and inverse hierarchies is substantial

\begin{equation} P_{\mu\mu}(NH)-P_{\mu\mu}(IH)\geq 0.30~~~~~~
P_{\bar\mu\bar\mu}(NH)-P_{\bar\mu\bar\mu}(IH)\geq 0.32
\end{equation}

\noindent and moreover it is also seen that the other two experiments (
No$\nu$a and LBNE) cannot offer such a possibility. Since this is the most favourable case for
hierarchy determination, we have also evaluated this probability difference using the second 
octant solution for $\theta_{23}$ (eq.(16)). The result is 

\begin{equation} P_{\mu\mu}(NH)-P_{\mu\mu}(IH)\geq 0.30~~~~~~
P_{\bar\mu\bar\mu}(NH)-P_{\bar\mu\bar\mu}(IH)\geq 0.29
\end{equation}

Furthermore if the hierarchy is normal, inspection of the bottom left panel of fig.8 shows that there 
are realistic prospects for $\delta_{CP}$ evaluation. In fact for LAGUNA, whose contour appears in
this scale similar to a circle, and for the $\mu e$ channel, we have $0.05\lesssim 
P_{\mu e}\lesssim 0.14$. In most of this interval, namely $0.06\lesssim P_{\mu e}\lesssim 0.13$, 
the two possible values of $P_{\bar\mu \bar e}$ are reasonably apart from each other: either 
$P_{\bar\mu \bar e}\lesssim 0.04$ or $0.17\lesssim P_{\bar\mu \bar e}\lesssim 0.21$. In this 
way the ambiguity in $\delta_{CP}$ can possibly be 
lifted. If otherwise $P_{\mu e}$ is found to lie close to either end of the interval, namely 0.05
or 0.14, the value of $\delta_{CP}$ is also unambigously determined, as there is only one 
possibility in each case: $\delta_{CP}\sim 135^o$ or $\delta_{CP}\sim 315^o$ respectively. So far
all results obtained are for the $\theta_{23}$ first octant solution (eq.(15)). Still for the
energy value being considered now (E=0.5 GeV) we show in fig.9 the same as fig.8 for the second 
octant solution ($\theta_{23}>$45$^{\circ}$, eq.(16)) whose results are similar. For larger energies
the distinction between the first and second octant solutions is even less clear as all contours 
pertaining to the three experiments become closer. We will return to this point in the next section.

\begin{table}[ht]
\centering
\begin{tabular}{|c|c|c|
}\hline
\multicolumn{3}{|c|}{810 km}  
\\ \hline
Channel &   E (GeV)  &   P \\ \hline 
$\mu\mu$        & 8.9$~-~$9.1   & 0.93$~-~$0.94 \\
$\bar\mu\bar e$ & 11.6$~-~$12.4 & $(1.75~-~2.0)\times 10^{-3}$ \\
$\bar\mu\bar\mu$ & 8.8$~-~$12.4 & 0.93$~-~$0.96 \\ \hline 
\end{tabular}
\caption{\it{The energy and oscillation probability for a
magic baseline in normal hierarchy at No$\nu$a (810 km): for each oscillation
channel and the neutrino energy range the probability shown is nearly independent of $\delta_{CP}$.
No magic baseline is found to exist for LBNE and LAGUNA in the energy range [0.5,12] GeV in normal
hierarchy.}}
\label{table1}
\end{table}

\begin{table}[ht]
\centering
\begin{tabular}{|c|c|c|c|c|} \hline
\multicolumn{3}{|c|}{810 km}  &  \multicolumn{2}{|c|}{2290 km} \\ \hline
Channel &   E (GeV)  &   P  &   E (GeV)  &  P  \\ \hline
$\mu e$        & 11.8$~-~$12.4   & $(1.1~-~1.2)\times 10^{-3}$               & ---  & ---  \\ 
$\mu \mu$ & 8.9$~-~$12.3 & 0.93$~-~$0.96 & --- & ---  \\ 
$\bar\mu\bar\mu$ & 8.8$~-~$8.9 & 0.92$~-~$0.93 & 4.7$~-~$5.0  & $(7.0~-~9.0)\times 10^{-2}$ \\ \hline
\end{tabular}
\caption{\it{The same as table 1 in inverse hierarchy for No$\nu$a (810 km) and LAGUNA (2290 km).
No magic baseline is found to exist for LBNE in the energy range [0.5,12] GeV in inverse hierarchy.}}
\label{table2}
\end{table}

Finally given the source-detector distances for No$\nu$a, LBNE and LAGUNA, we have estimated the
necessary neutrino energy for each distance to become a magic baseline and the corresponding oscillation 
probability. We analysed the energy interval [0.5,12] GeV for which all three detectors are designed. 
Our results are shown in tables 1 and 2. For each oscillation channel we show the energy range where
the probability is nearly constant with detectors located at 810 km from the neutrino source
(No$\nu$a) and 2290 km (LAGUNA). For LBNE no magic baseline is found to exist in this energy range.
We note that for the $\bar\mu\bar\mu$ channel both distances can become magic baselines, whereas for 
$\mu e$ with normal and $\bar\mu \bar e$ with inverse hierarchy none of the distances is suitable.
Since at the magic baseline the oscillation probability becomes $\delta_{CP}$ independent, any 
significant deviation from the oscillation prediction will be the signature of nonstandard interactions 
\cite{Gago:2009ij},\cite{Das:2010sd},\cite{Coelho:2012bp}.

\section{T2K and No$\nu$a}

The main objective of the T2K experiment \cite {Abe:2011ks} is to discover $\nu_{e}$ appearance from 
$\nu_{\mu}$. The collaboration reported their first results in June 2011 \cite{Abe:2011sj} which were 
later improved \cite{Sakashita} and also reported evidence for $\nu_{\mu}$ disappearance \cite{Abe:2012gx}. 
The experiment is a long baseline off-axis one with a 295 km source-detector distance from Tokai (J-PARC) 
to Kamioka (SuperKamiokande) and its peak neutrino energy is around 600 MeV.

Our aim in this section is twofold: to obtain the predictions for the oscillation probabilities $P_{\mu e}$ 
and $P_{\mu\mu}$ at T2K displaying them in terms of biprobability plots and to explore the possible
complementarity between T2K and No$\nu$a. Our results are shown in the four panels of fig.10. We
consider normal and inverse hierarchies (left and right panels) and first and second octant solutions 
for $\theta_{23}$ (top and bottom panels).

We recall that in the previous section the prospective data from one experiment were compared at
the same energy for different hierarchies and/or different octants. In no case the contours from
two different experiments were ever directly compared. Thus it was ensured that the comparisons were
made at the same point in the oscillation phase. Exploring the complementarity between T2K and
No$\nu$a, since the two experiments operate at different phases, requires a reduction of the
probability to the same phase. Given the fact that T2K runs with a peak neutrino energy E=0.6 GeV 
at 295 km distance one must operate No$\nu$a, whose source-detector distance is 810 km, at a
neutrino energy value satisfying
\begin{equation}
\left(\frac{L}{E}\right)_{No{\nu}a}=\left(\frac{L}{E}\right)_{T2K}
\end{equation} 
which gives ${\rm E}_{No\nu a}\simeq$ 1.65 GeV. In fig.10 the dotted contours refer to T2K (E=0.6 GeV) 
and the full contours to No$\nu$a (E=1.65 GeV). Hence observing the left and right panels of fig.10
on each row, it is seen that
$$(P_{\mu e})_{No{\nu}a}>(P_{\mu e})_{T2K}~~({\rm normal~hierarchy})~~~~
(P_{\mu e})_{No{\nu}a}<(P_{\mu e})_{T2K}~~({\rm inverse~hierarchy})~.$$
On the other hand observing the top and bottom panels on each column,
$$(P_{\mu \mu})_{No{\nu}a}>(P_{\mu \mu})_{T2K}~~(\theta_{23}<45^{\circ})~~~~
(P_{\mu \mu})_{No{\nu}a}<(P_{\mu \mu})_{T2K}~~(\theta_{23}>45^{\circ})~.$$
Although these inequalities apply for any value of $\delta_{CP}$, the differences in the probabilities 
are so small that the possibility of ever detecting them in this way is slim.

\section{Summary and conclusions}

We have investigated the prospects for distinguishing normal from inverse neutrino mass hierarchies,
$\delta_{CP}$ determination and first vs. second octant $\theta_{23}$ solutions with T2K, No$\nu$a, 
LBNE and LAGUNA long baseline experiments. We examined the oscillation channels which are possibly 
relevant for these, namely $\mu e$, $\mu\mu$ and their antineutrino counterparts. Owing to the 
baseline distances involved, the neutrinos are assumed to pass through the Earth's mantle only. 
The starting point for our numerical analysis is the general formula for the matter oscillation 
probability derived in section 2. The discrepancies between the results from the leading term
approximations existing in the literature based on the smallness of the mass square differences 
ratio or the $\theta_{13}$ mixing angle were also examined. 

The main results of our paper are described in section 3.2 and displayed in fig.8. They are 
suggestive of the importance of LAGUNA and its operation at a neutrino energy around 0.5 GeV. 
Otherwise we have found that there are also good possibilities to distinguish between hierarchies with
other experiments provided the neutrino energy can be tuned to reasonable accuracy. To this end the 
oscillation channels which can provide the best information are the muon and antimuon disappearance
ones, namely $\nu_{\mu}\rightarrow \nu_{\mu}$ and $\bar\nu_{\mu}\rightarrow \bar\nu_{\mu}$. For
a neutrino energy E $\simeq$ 1 GeV the LBNE far detector will be confronted with a probability 
$P_{\bar\mu\bar\mu}\simeq 0.48$ for inverse or a probability $0.62\lesssim P_{\bar\mu\bar\mu}
\lesssim 0.66$ for normal hierarchy (see fig.3, bottom panels), whereas for No$\nu$a these are 
$0.65\lesssim P_{\bar\mu\bar\mu}\lesssim 0.68$ (inverse) and $0.55\lesssim P_{\bar\mu\bar\mu}
\lesssim 0.56$ (normal) as can also be seen from fig.3 (top panels). On the other hand for a 
neutrino energy E $\simeq$ 0.5 GeV the possibility of hierarchy differentiation looks much brighter. 
In fact if the hierarchy is inverse, for the $\mu\mu$ channel in the LAGUNA far detector one 
expects $0.03\lesssim P_{\mu\mu}\lesssim 0.07$, if it is normal then $0.37\lesssim P_{\mu\mu}
\lesssim 0.40$ (fig.8, top panels). Moreover for the $\bar\mu\bar\mu$ channel and for LAGUNA 
with inverse hierarchy we obtain $0.03\lesssim P_{\bar\mu\bar\mu}\lesssim 0.06$ and with normal 
hierarchy $0.36\lesssim P_{\bar\mu\bar\mu}\lesssim 0.50$ (fig.8, middle panels). The difference 
between the oscillation probabilities $P_{\bar\mu\bar\mu}$ for normal and inverse hierarchy is 
thus 0.32 or larger and for $P_{\mu\mu}$ it is 0.30 or larger (see eq.(22)). Hence the ${\mu\mu}$ 
and ${\bar\mu\bar\mu}$ channels at LAGUNA with a neutrino energy E $\simeq$ 0.5 GeV seem to be the 
most promising ones to explore in order to find out the mass hierarchy. On the other hand the
complementarity between T2K and No$\nu$a, presented in fig.10, does not offer a clear perspective
for the mass hierarchy determination.

For the $\theta_{23}$ solution in the second octant, we have also checked the possible distinction
between hierarchies. The chances look almost identical as for the first octant solution, although 
slightly disfavoured in the ${\bar\mu\bar\mu}$ channel (see eq.(23)). Distinguishing between
first and second octant solutions on the other hand looks difficult as can be seen from the
comparison between figs.8 and 9 and from fig.10. 

As for $\delta_{CP}$ determination the LAGUNA far detector offers good chances for its 
feasability on the basis of observing the $\mu e$ and $\bar\mu \bar e$ channels. Again the neutrino 
energy must be tuned to E $\simeq$ 0.5 GeV. Then if one is able to experimentally distinguish between 
probabilities $P_{\bar\mu \bar e}\lesssim 0.04$ and $0.17\lesssim P_{\bar\mu \bar e}\lesssim 0.21$, 
a relatively narrow interval for $\delta_{CP}$ can be determined (see fig.8, bottom left panel).
Again, for $\theta_{23}$ in the second octant the prospects are practically the same.

Finally, regarding nonstandard interactions, since the baseline distances are a priori fixed,
these will become magic for a conveniently chosen value of the neutrino energy. This energy was 
evaluated in section 3 along with the corresponding oscillation probability. Given this 
probability, any deviation from such a value is a signature of nonstandard interactions. In the 
appropriate neutrino energy interval for the three experiments, namely [0.5,12] GeV, the range 
to search for magic baselines is ${\rm E}_{\nu}\geq$ 8.8 GeV for No$\nu$a and (4.7$~-~$5.0) GeV 
for LAGUNA. No magic baseline apparently exists for LBNE in the above interval. Detailed results 
are shown in tables 1 and 2.

To conclude, accelerator based long baseline neutrino experiments offer good prospects to 
discover the $\delta_{CP}$ phase range, the mass hierarchy and may be nonstandard interactions
if they really exist.

\section*{Acknowledgments}
We acknowledge discussions with Evgeni Akhmedov and Luis Lavoura. C. R. Das gratefully acknowledges 
a scholarship from Funda\c{c}\~{a}o para a Ci\^{e}ncia e a Tecnologia (FCT, Portugal) 
ref. SFRH/BPD/41091/2007. This work was partially supported by FCT through the projects 
CERN/FP/123580/2011 PTDC/FIS/ 098188/2008 and CFTP-FCT Unit 777 which are partially funded
through POCTI (FEDER).


\begin{figure}
\centering
\vspace{-1.2cm}
\hspace*{-2.5cm}
\includegraphics[height=185mm,keepaspectratio=true,angle=0]{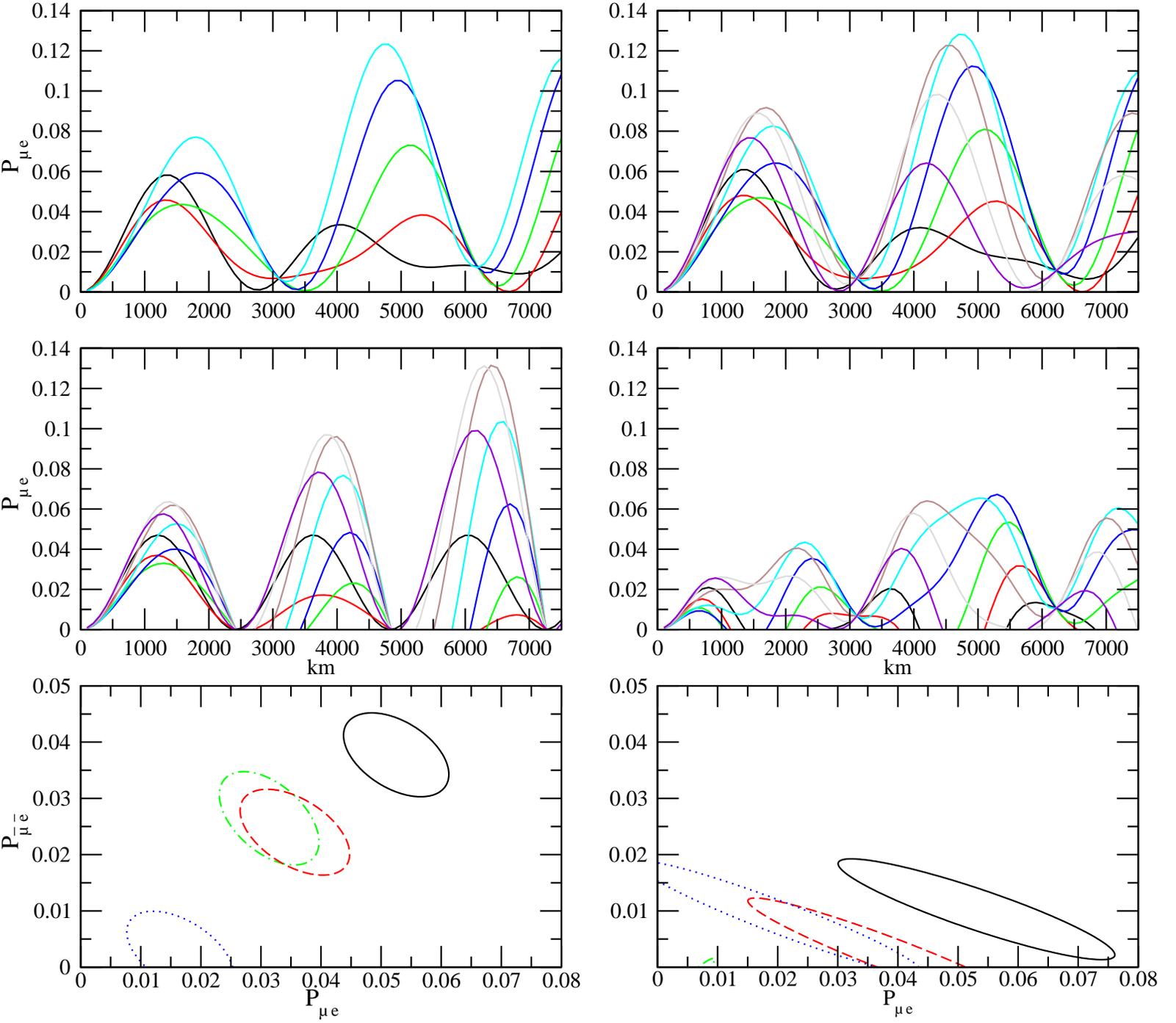}
\caption{ \it Four top panels (clockwise): $P_{\mu e}$ as a function of distance from 0 to
7500 km evaluated with our numerical approach, and with the approximate approximations given
in \cite{Akhmedov:2004ny},\cite{Cervera:2000kp},\cite{Minakata:2001qm} for eight values of 
$\delta_{CP}$ equally spaced from 0$^0$ to 360$^0$. In our case (top left panel) constant 
$\delta_{CP}$ lines for 225$^0$, 270$^0$, 315$^0$ coincide with those for 135$^0$, 
90$^0$, 45$^0$. Two bottom panels: biprobability plots $P_{\mu e}, P_{\bar\mu \bar e}$ at 
810 km (left) and 2290 km (right) from the neutrino source. Full, dotted, dashed and 
dot-dashed contours are obtained from our approach, and from the approximate expressions 
in refs. \cite{Akhmedov:2004ny},\cite{Cervera:2000kp},\cite{Minakata:2001qm}. All results 
are for $\theta_{23}$ in the first octant (eq.(15)), normal hierarchy and neutrino energy 2.3 GeV.}
\label{fig1}
\end{figure}

\begin{figure}
\centering
\vspace{-3.2cm}
\hspace*{-2.4cm}
\includegraphics[height=200mm,width=200mm,angle=0]{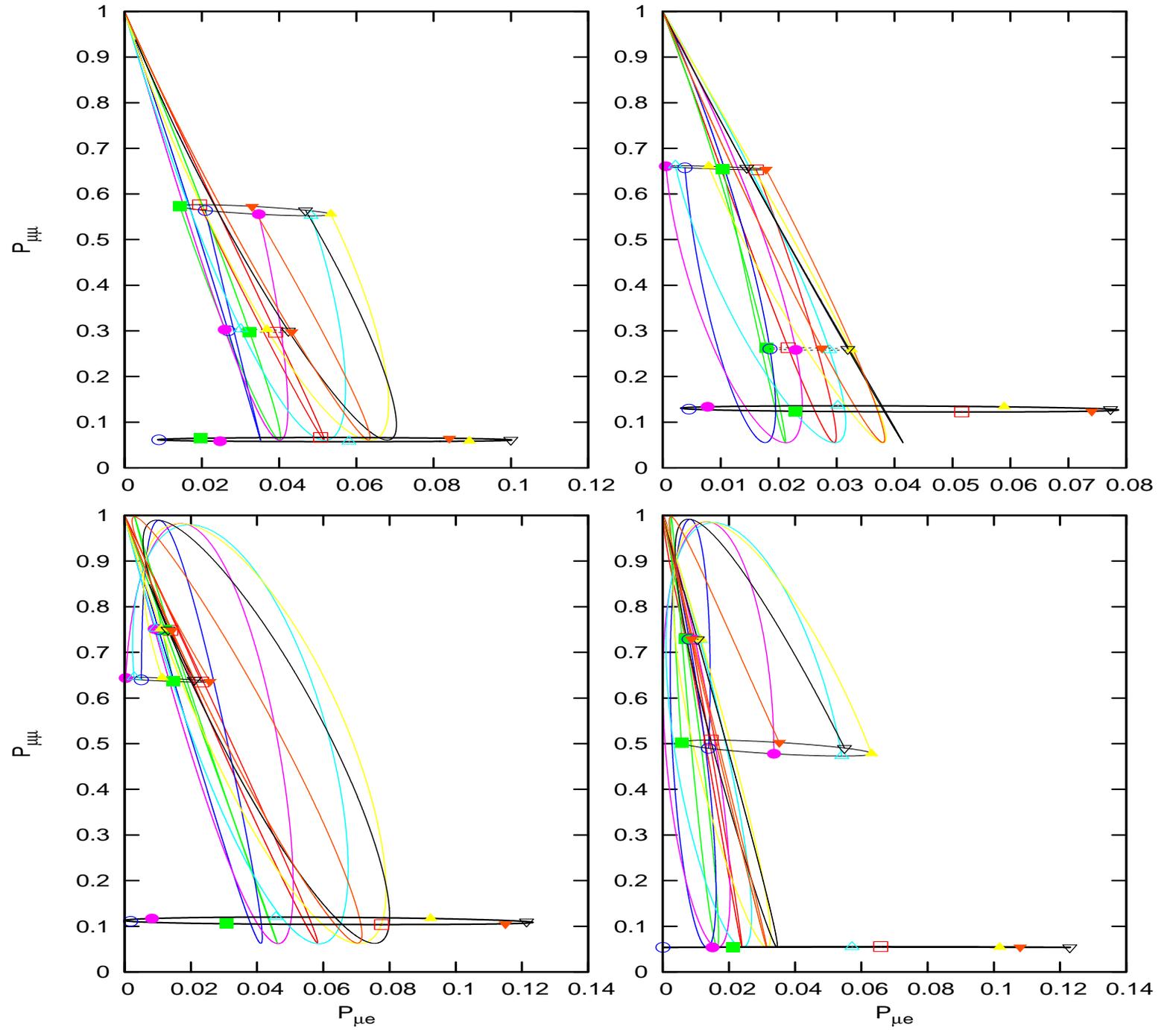}
\vspace{-1.2cm}
\caption{ \it Biprobability plots for the ($\mu e,\mu\mu$) channel pair with energy in 
the range [0.5,100] GeV and $\theta_{23}<$45$^\circ$ (eq.(15)) showing the constant $\delta_{CP}$ 
curves (which merge at coordinates (0,1) for large energy) and the constant energy contours. 
Top panels: 810 km distance with (from top to bottom) 1 GeV, 2.3 GeV, 0.5 GeV contours. 
Bottom panels: 1290 km distance with (from top to bottom) 7 GeV, 1 GeV. 0.5 GeV
contours. Left and right panels: normal and inverse hierarchy respectively.
Points marked as $\boxdot,\blacksquare,\odot,\bullet,
\vartriangle,\blacktriangle,\triangledown,\blacktriangledown$ are for $\delta_{CP}=$
0$^0$, 45$^0$, 90$^0$, 135$^0$, 180$^0$, 225$^0$, 270$^0$, 315$^0$.}
\label{fig2}
\end{figure}

\begin{figure}
\centering
\vspace{-2.6cm}
\hspace*{-2.4cm}
\includegraphics[height=235mm,width=200mm,keepaspectratio=false,angle=0]{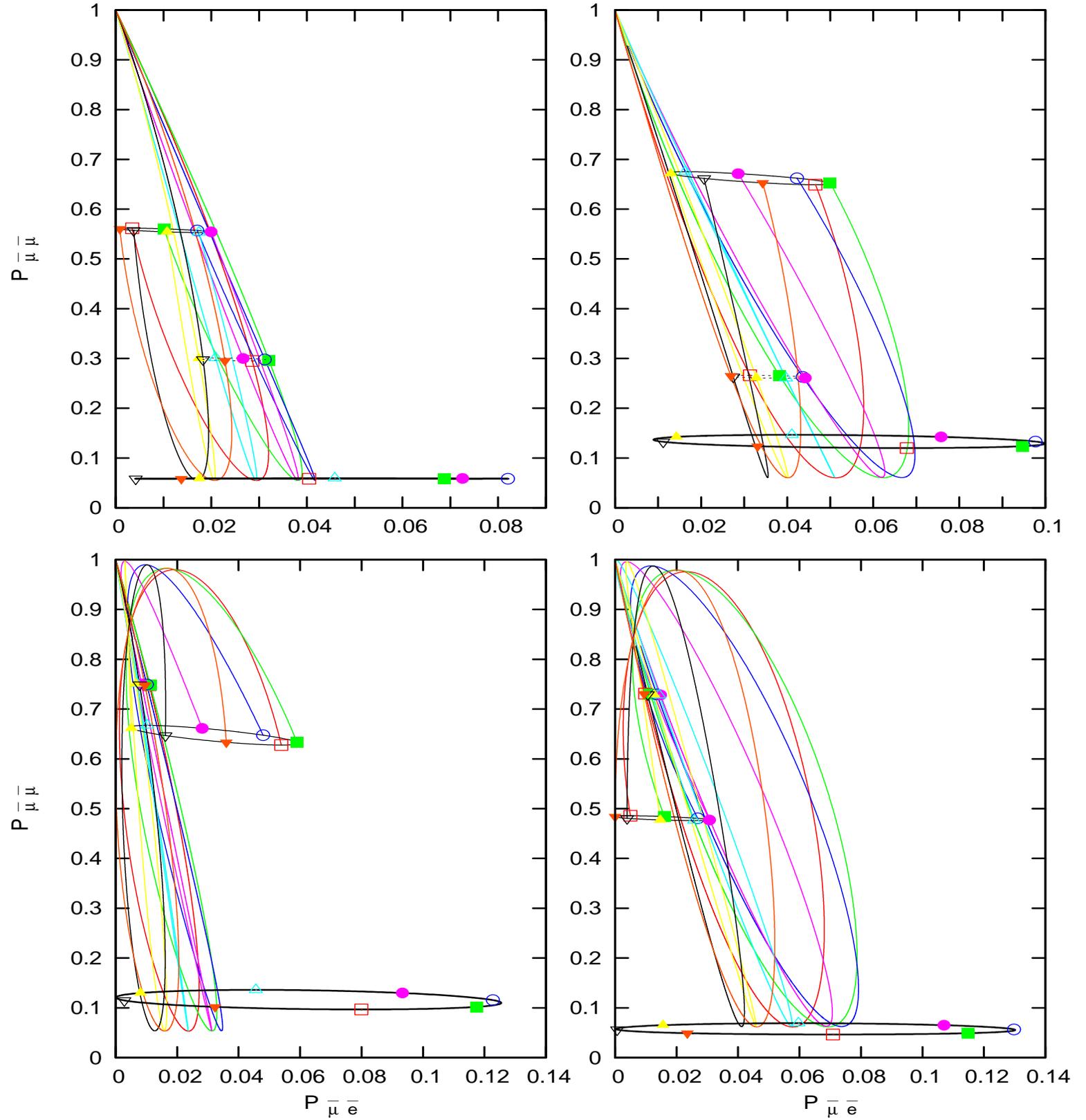}
\vspace{-1.0cm}
\caption{ \it The same as fig.\ref{fig2} for the ($\bar\mu \bar e,\bar\mu\bar\mu$) channel pair.}
\label{fig3}
\end{figure}

\begin{figure}
\centering
\vspace{-2.5cm}
\hspace*{-2.4cm}
\includegraphics[height=215mm,width=200mm,keepaspectratio=false,angle=0]{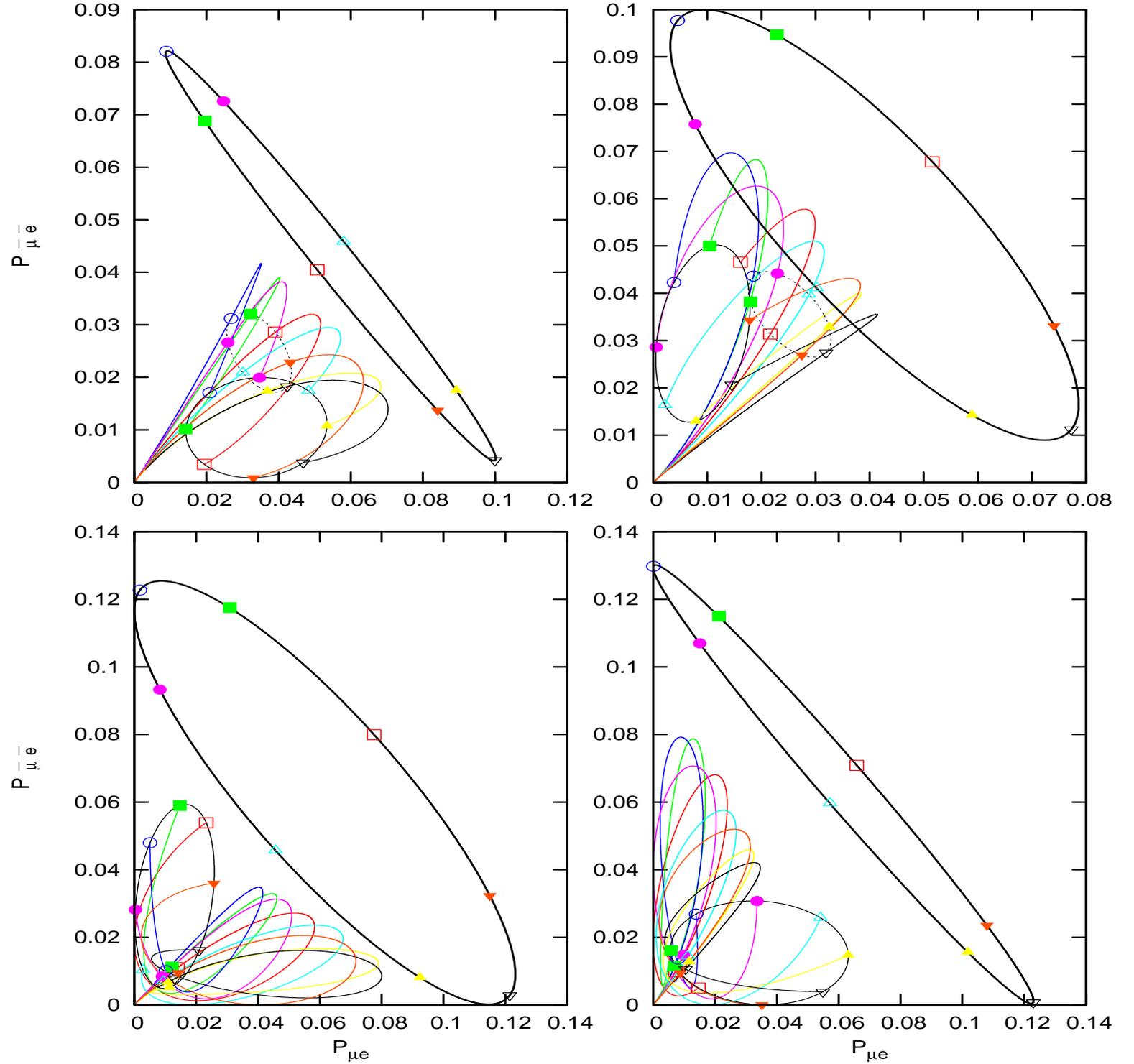}
\vspace{-1.2cm}
\caption{ \it Biprobability plot for the ($\mu e,\bar\mu\bar e$) channel pair with energy in 
the range [0.5,100] GeV, $\delta_{CP}<$45$^{\circ}$. Top panels:
the contours for 810 km with 2.3 GeV (dotted), 1 GeV (thin full) 0.5 GeV (thick full). Bottom
panels: the contours for 1290 km with 7 GeV (dotted), 1 GeV (thin full) 0.5 GeV (thick full). 
Due to the figure scale, the contour for 7 GeV in the bottom panel can hardly be seen. 
The constant $\delta_{CP}$ curves merge for large energy at coordinates (0,0) and the 
values of $\delta_{CP}$ are marked as in figs.2 and 3.}
\label{fig4}
\end{figure}

\begin{figure}
\centering
\vspace{-2.2cm}
\hspace*{-2.4cm}
\includegraphics[height=195mm,width=205mm,keepaspectratio=false,angle=0]{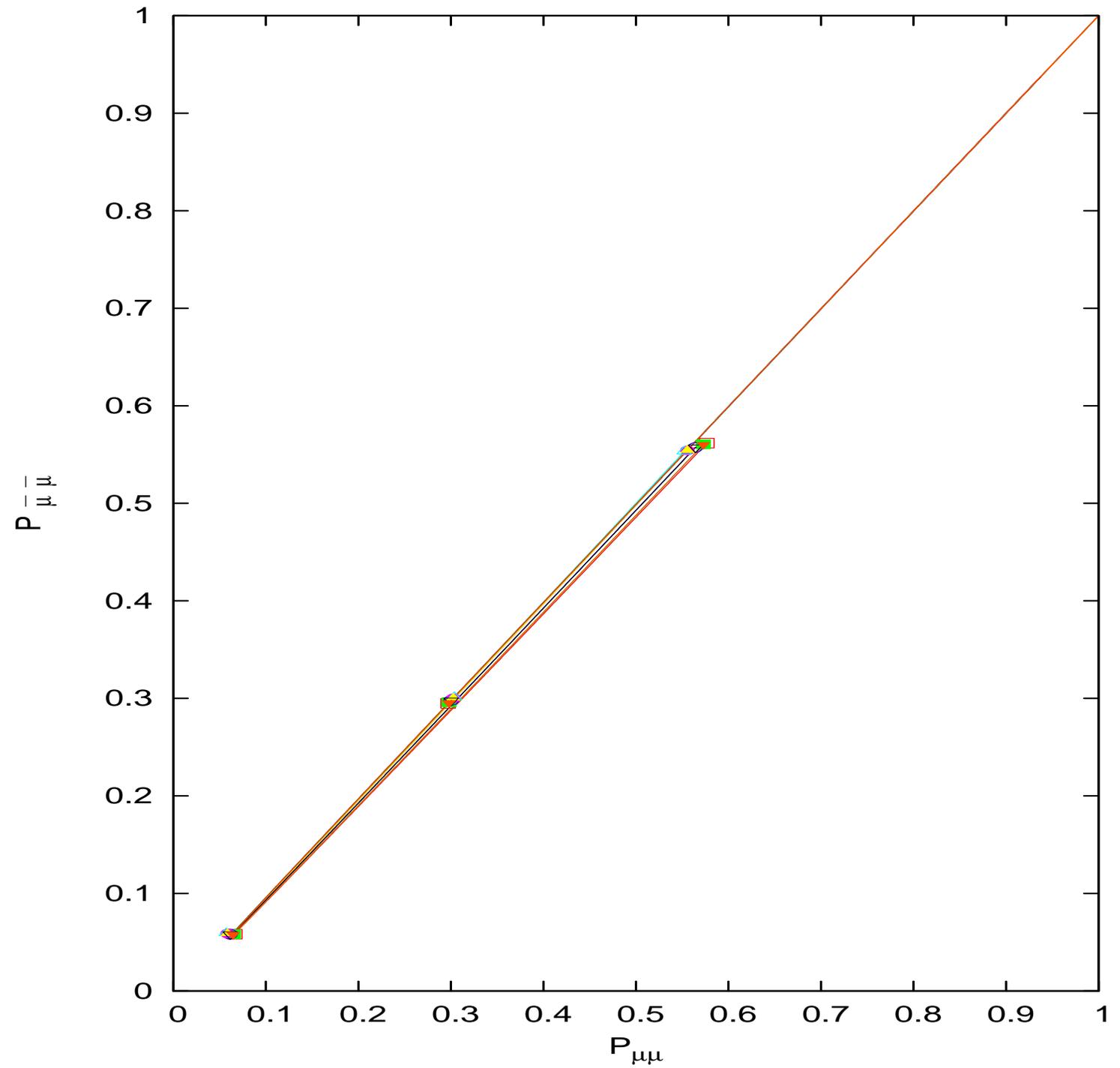}
\vspace{-0.5cm}
\caption{ \it Biprobability curves for constant $\delta_{CP}$ for the ($\mu\mu,
\bar\mu\bar\mu$) channel pair at 810 km source/detector distance in normal hierarchy with
$\theta_{23}<$45$^{\circ}$. Their intersections with the constant energy contours are 
marked. Their closeness prevents determination of $\delta_{CP}$. The neutrino energy 
interval is [0.5,12] GeV.}
\label{fig5}
\end{figure}

\begin{figure}
\centering
\vspace{-1.2cm}
\hspace*{-2.5cm}
\includegraphics[height=205mm,width=205mm,angle=0]{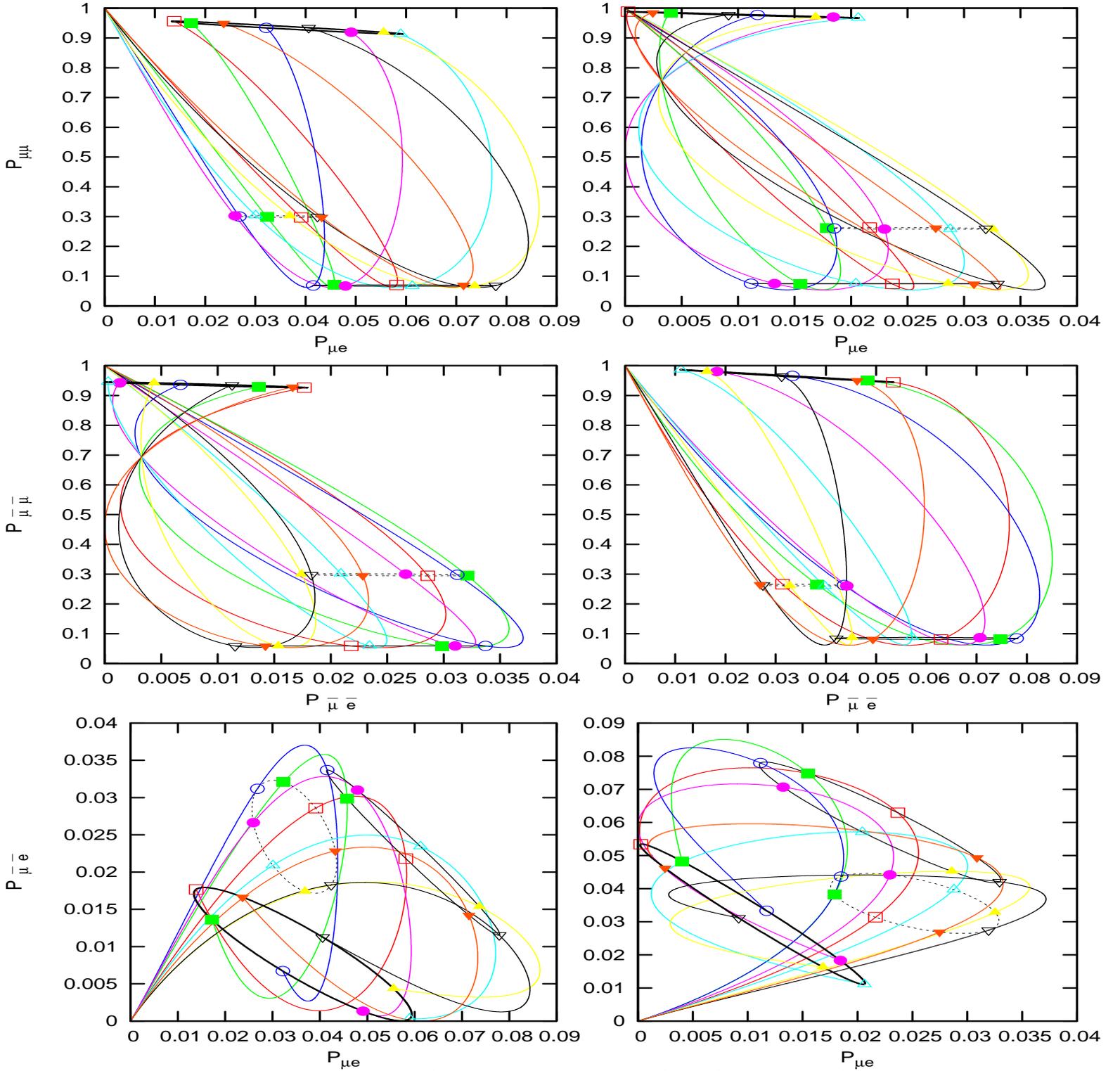}
\vspace{-1.3cm}
\caption{ \it Biprobability graphs for neutrino energy 2.3 GeV ($\theta_{23}<$45$^\circ$,
eq.(15)): the curves for constant $\delta_{CP}$ for varying distance and the contours 
for No$\nu$a (dotted), LBNE (thin full) and LAGUNA (thick full). Left and 
right panels are for normal and inverse hierarchy. Top, middle and bottom panels: ($\mu e,\mu\mu$),
($\bar\mu \bar e,\bar\mu\bar\mu$) and ($\mu e,\bar\mu \bar e$) channel pairs. Values of $\delta_{CP}$ 
on each contour are marked as in fig.2:
$\boxdot,\blacksquare,\odot,\bullet,\vartriangle,\blacktriangle,\triangledown,\blacktriangledown$
are for $\delta_{CP}=$0$^0$, 45$^0$, 90$^0$, 135$^0$, 180$^0$, 225$^0$, 270$^0$, 315$^0$.
Merging of $\delta_{CP}$ curves occurs for 0 km distance.}
\label{fig6}
\end{figure}

\begin{figure}[htb]
\centering
\hspace*{-1.8cm}
\includegraphics[height=185mm,width=185mm,keepaspectratio=true,angle=0]{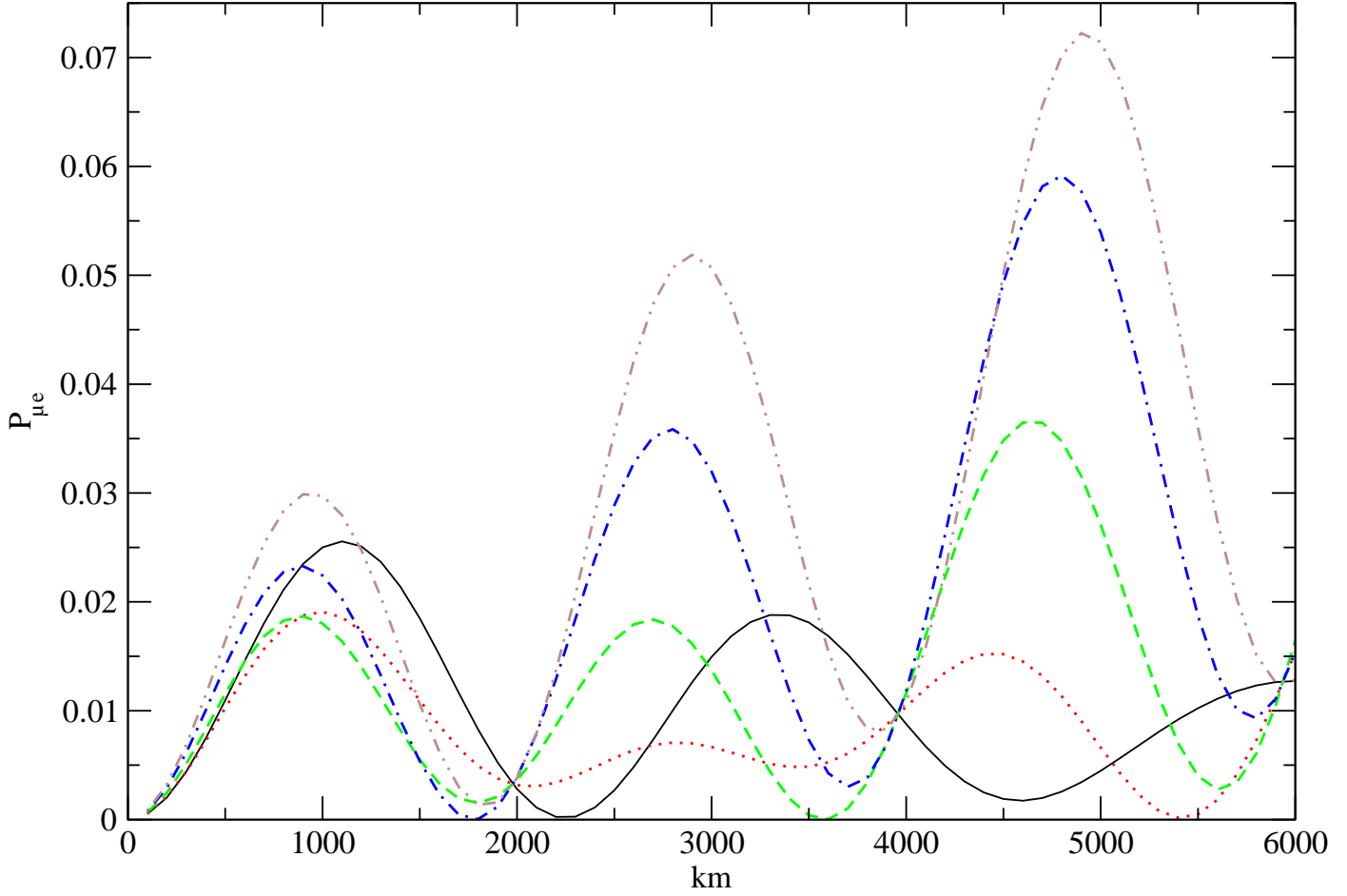}
\caption{ \it $P_{\mu e}$ (inverse hierarchy, $\theta_{23}<$45$^{\circ}$, eq.(15))) as a function 
of distance from 0 to 6000 km for eight values of $\delta_{CP}$ equally spaced from 0$^0$ to 360$^0$. 
Full, dotted, dashed, dot-dashed and dot-double 
dashed lines are for $\delta_{CP}$=0$^0$, 45$^0$, 90$^0$, 135$^0$, 180$^0$ respectively. $\delta_{CP}$=
225$^0$, 270$^0$, 315$^0$ lines coincide with 135$^0$, 90$^0$, 45$^0$. The neutrino energy is 2.3 GeV.
The analogue of this graph for normal hierarchy is shown in the top left panel of fig.1.}  
\label{fig7}
\end{figure}

\begin{figure}
\centering
\vspace{-1.0cm}
\hspace*{-2.4cm}
\includegraphics[height=200mm,width=205mm,angle=0]{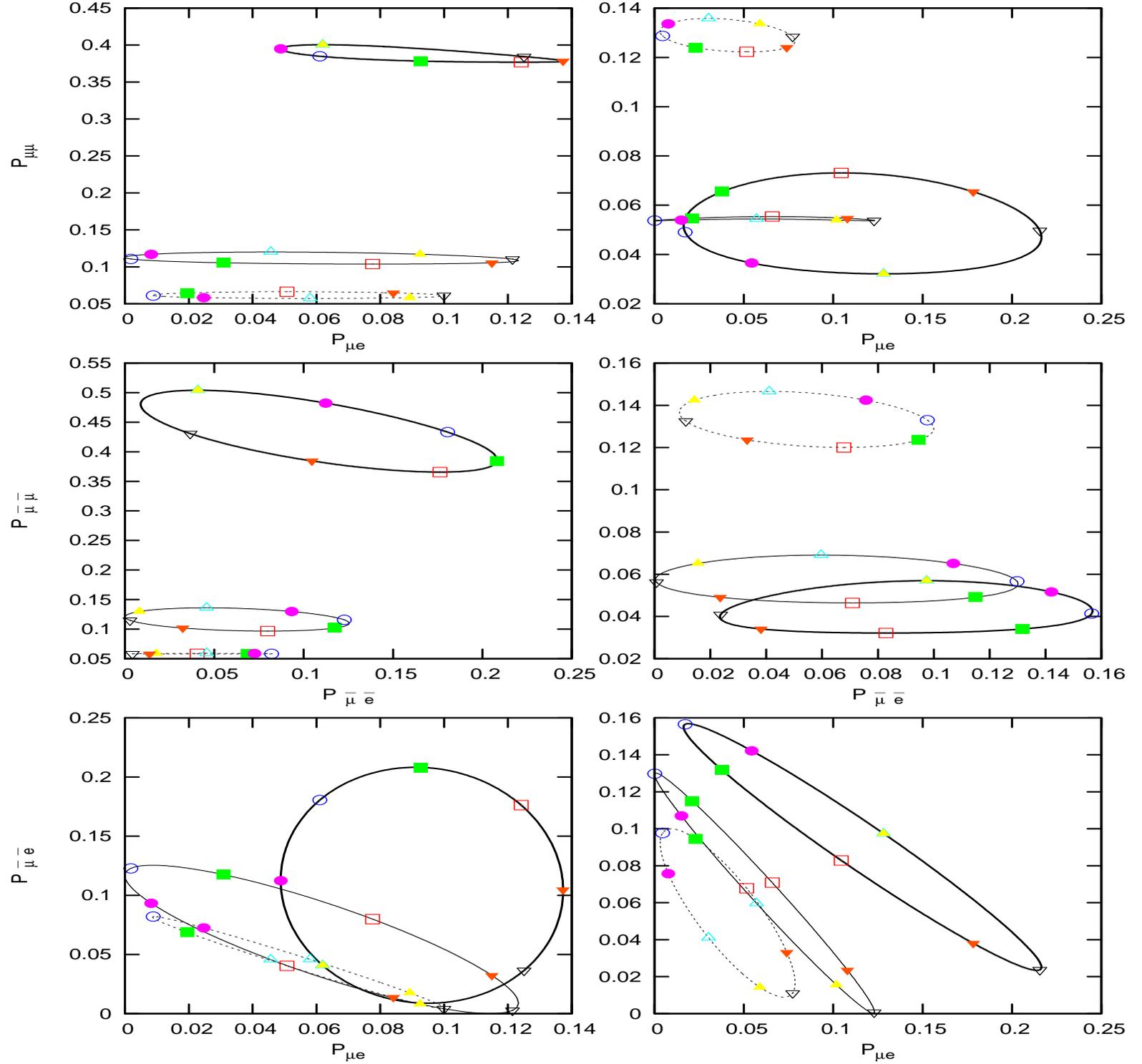}
\vspace{-1.2cm}
\caption{ \it Biprobability graphs for neutrino energy 0.5 GeV ($\theta_{23}<$45$^{\circ}$, 
eq.(15)): the contours for No$\nu$a (dotted), LBNE (dashed) and LAGUNA (full). The values 
of $\delta_{CP}$ on each contour are marked as in fig.2:  
$\boxdot,\blacksquare,\odot,\bullet,\vartriangle,\blacktriangle,\triangledown,\blacktriangledown$ 
are for $\delta_{CP}=$0$^0$, 45$^0$, 90$^0$, 135$^0$, 180$^0$, 225$^0$, 270$^0$, 315$^0$.
Left and right panels: normal and inverse hierarchy. Top, middle and bottom panels: ($\mu e,\mu\mu$), 
($\bar\mu \bar e,\bar\mu\bar\mu$) and ($\mu e,\bar\mu \bar e$) channel pairs. The determination of the 
mass hierarchy and $\delta_{CP}$ range appear more favourable in this case than for 2.3 GeV 
(fig.\ref{fig6}).}
\label{fig8}
\end{figure}

\begin{figure}
\centering
\vspace{-1.8cm}
\hspace*{-2.4cm}
\includegraphics[height=200mm,width=205mm,angle=0]{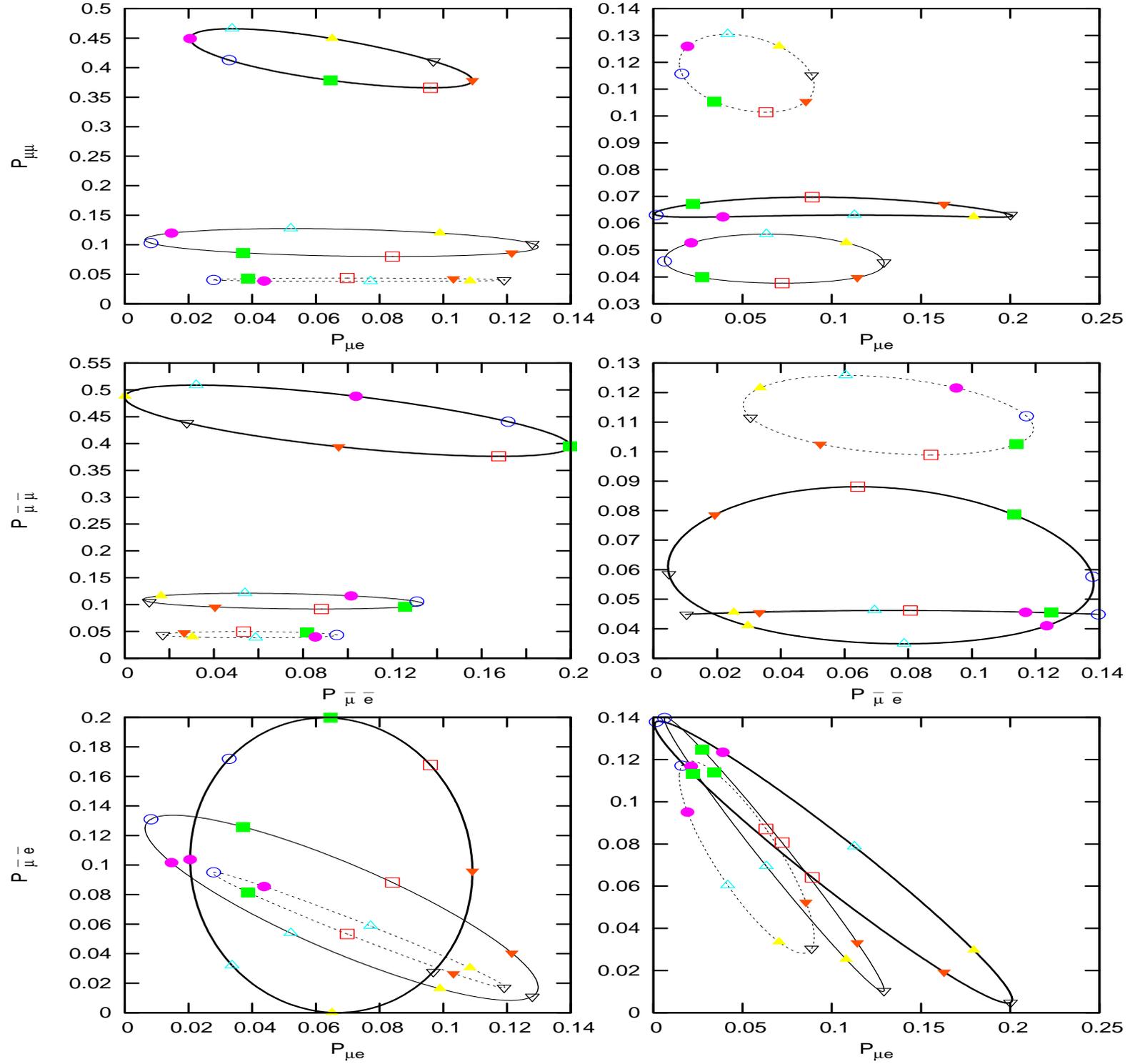}
\vspace{-0.8cm}
\caption{ \it The same as fig.8 for the second octant solution ($\theta_{23}>45^\circ$, eq.(16)).
From comparison with fig.8 it is seen that hardly any distinction can be made between the two
solutions.}
\label{fig9}
\end{figure}

\begin{figure}
\centering
\vspace{-1.8cm}
\hspace*{-2.4cm}
\includegraphics[height=210mm,width=202mm,angle=0]{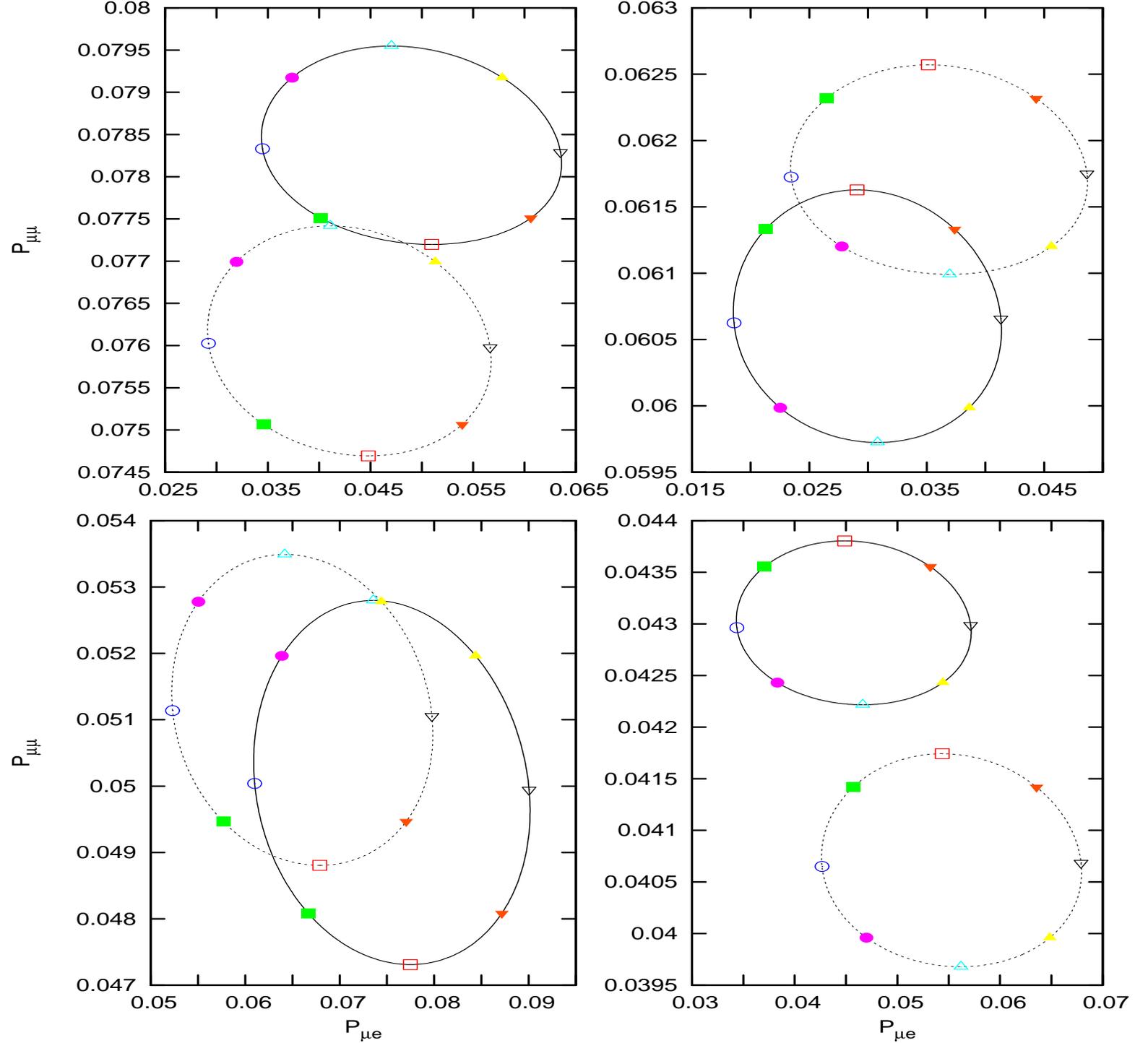}
\vspace{-0.8cm}
\caption{ \it Comparison between T2K and No$\nu$a: left and right panels are for normal and
inverse hierarchy, top and bottom ones for $\theta_{23}$ in the first (eq.(15)) and second octant
(eq.(16)). Dotted contours are for T2K and full contours for No$\nu$a. Values of $\delta_{CP}$ 
are marked in each contour and follow the conventions of the previous figures.}
\label{fig10}
\end{figure}

\end{document}